\documentclass[aps, reprint, preprintnumbers, footinbib, showpacs, citeautoscript, superscriptaddress, flushbottom, balancelastpage]{revtex4-1}

\usepackage[T1]{fontenc}
\usepackage[english]{babel}

\usepackage{amsmath}

\usepackage{times}
\usepackage{textcomp}

\usepackage{bm}

\usepackage{graphicx}
\usepackage{xcolor}
\usepackage[colorlinks, urlcolor=blue, citecolor=blue, linkcolor=blue, pdfstartview=FitH]{hyperref}
\usepackage[all]{hypcap}

\newcommand{\e}{\mathrm{e}}
\renewcommand{\d}{\mathrm{d}}
\renewcommand{\i}{{\rm i}}

\newcommand{\aver}[1]{\left\langle #1\right\rangle}
\newcommand{\sign}{\mathop{\rm sign}}

\newcommand\affiSOLAB{Spin\ Optics\ Laboratory, Saint~Petersburg\ State\ University, 198504\ Peterhof, St.~Petersburg, Russia}
\newcommand\affiIOFFE{Ioffe\ Institute, 194021\ St.~Petersburg, Russia}
\newcommand\affiSOI{S.~I.\ Vavilov\ State\ Optical\ Institute, 199053\ St.~Petersburg, Russia}

\begin{document}

\title{Homogenization of Doppler broadening in spin noise spectroscopy}

\author{M.~Yu.~Petrov}
\affiliation{\affiSOLAB}

\author{I.~I.~Ryzhov}
\affiliation{\affiSOLAB}

\author{D.~S.~Smirnov}
\affiliation{\affiIOFFE}

\author{L.~Yu.~Belyaev}
\affiliation{\affiSOLAB}

\author{R.~A.~Potekhin}
\affiliation{\affiSOLAB}

\author{M.~M.~Glazov}
\affiliation{\affiIOFFE}
\affiliation{\affiSOLAB}

\author{V.~N.~Kulyasov}
\affiliation{\affiSOI}

\author{G.~G.~Kozlov}
\affiliation{\affiSOLAB}

\author{E.~B.~Aleksandrov}
\affiliation{\affiIOFFE}
\affiliation{\affiSOI}

\author{V.~S.~Zapasskii}
\affiliation{\affiSOLAB}


\date{\today}

\begin{abstract}
The spin noise spectroscopy, being a nonperturbative linear optics tool, is still reputed to reveal a number of capabilities specific to nonlinear optics techniques. The new effect of the Doppler broadening homogenization  discovered in this work essentially widens these unique properties of spin noise spectroscopy. We investigate spin noise of a classical system---cesium atoms vapor with admixture of buffer gas---by measuring the spin-induced Faraday rotation fluctuations in the region of D${_2}$ line. The line, under our experimental conditions, is strongly inhomogeneously broadened due to the Doppler effect. Despite that, optical spectrum of the spin noise power has the shape typical for the homogeneously broadened line with a dip at the line center. This fact is in stark contrast with the results of previous studies of inhomogeneous quantum dot ensembles and Doppler broadened atomic systems. In addition, the two-color spin-noise measurements have shown, in a highly spectacular way, that fluctuations of the Faraday rotation within the line are either correlated or anticorrelated depending on  whether the two wavelengths lie on the same side, or on different sides of the resonance. The experimental data are interpreted in the frame of the developed theoretical model which takes into account both kinetics and spin dynamics of Cs atoms. It is shown that the unexpected behavior of the Faraday rotation noise spectra and effective homogenization of the optical transition in the spin-noise measurements are related to smallness of the momentum relaxation time of the atoms as compared with their spin relaxation time. Our findings demonstrate novel abilities of spin noise spectroscopy for studying dynamic properties of inhomogeneously broadened ensembles of randomly moving spins.

\end{abstract}

\maketitle

\section{Introduction}

The idea of detecting magnetic resonance in the Faraday-rotation (FR) noise spectrum, first realized in 1981~\cite{aleksandrov81}, has gained nowadays a considerable popularity, see, e.g., reviews~\cite{Zapasskii:13,Oestreich-review,SinitsynReview}.
Primarily, the main advantage of the spin noise spectroscopy (SNS) was recognized to be its nonperturbative character: A weak light beam probing the sample in the region of its transparency is unable to produce any real transitions and leaves the system intact. 
At present, however, it became clear that specific merits of the SNS extend far beyond its nonperturbativity~\cite{NonlinearSNS}. 
In the framework of linear optics, this technique allows one, along with getting the data traditionally provided by the ESR spectroscopy, such as Land\'{e} factors and spin relaxation rates~\cite{eh_noise}, to obtain information about spatial characteristics of the spin system~\cite{Z-scan}, mechanisms of broadening of optical transitions~\cite{Zapasskii13}, kinetic parameters of motion of the spin carriers~\cite{pershin-two-beam-sns,PhysRevB.92.045308,2017arXiv170709832L}, dynamics of local magnetic fields in the sample~\cite{OpticalField}, etc. 

Additional informative abilities of the SNS arise under conditions of strong or resonant optical ``probing'', when the effects of nonlinear optics become essential, and the SNS is getting fundamentally perturbative~\cite{noise-trions}. 
In these cases, the SNS provides novel information about mechanisms of spin-photon interaction as well as about the nonequilibrium dynamics of charge carriers and spins~\cite{Nonresonant_nonequilibrium, Glazov_Keldysh}.

Sensitivity of the SNS is determined by efficiency of conversion of the spin-system magnetization into the FR. 
This efficiency, in turn, is controlled by detuning of the probe beam wavelength with respect to the relevant optical transition. 
This is the reason why ensembles of free paramagnetic atoms, with narrow allowed optical transitions and great efficiency of the above conversion, became the first objects of the SNS~\cite{aleksandrov81,Crooker_Noise}. 
It should be noted, however, that the resonant absorption lines, in atomic vapors, are usually strongly inhomogeneously broadened due to the Doppler effect, with the linewidth of individual atoms being much smaller than the total width of the Doppler-broadened transition. 
Since every atom of the ensemble exhibits frequent collisions accompanied by changes of its velocity, its resonant frequency is constantly bouncing  over the Doppler-broadened profile of the line, providing certain fluctuations of the detuning. 
This effect was expected to be especially pronounced for the probe beam wavelength lying in close vicinity of (or inside) the absorption line where the main contribution to the FR noise is made by atoms with smallest detunings on the order of homogeneous linewidth. 

Previously, the effects of inhomogeneous broadening on the optical FR noise spectra have been studied for semiconductor quantum dots, where the broadening results from static fluctuations of the quantum dot resonance frequencies caused by variation of their geometry and composition~\cite{gi2012noise,Zapasskii13,Yang:2014aa}. 
These works have shown that (i) the spin noise (SN) signal is the greatest for the probe beam tuned at maximum of the absorption band~\cite{Zapasskii13} and (ii) in the two-color experiment, where two linearly polarized probe beams with different wavelengths are used, the correlations between the fluctuating FR of the beams are absent unless the detuning between the beams becomes comparable with the homogeneous linewidth of the optical transition~\cite{Yang:2014aa}. In recent work \cite{Ma:2017aa} the two limiting cases of homogeneous and inhomogeneous broadening have been realized in Rb gas vapors.

The goal of this paper is to examine spectral behavior of the spin noise resonance of alkali atoms throughout a resonance optical transition. 
Here, we demonstrate experimentally and theoretically that motion of atoms reveals itself, in the spin noise spectra, in a very peculiar way leading to effective homogenization of the Doppler broadening. 
Particularly, the FR noise vanishes for the probe beam tuned to the absorption line center. 
Moreover, in the two-color experiment the pronounced correlations in the FR noise of the two beams are present when both  beams are tuned to one side of the resonance, while, for the beams tuned to opposite sides of the resonance, fluctuations of the FR turn out to be anticorrelated. 
 Experimental results are supported by a microscopic model of the FR fluctuations in the atomic gas.

The paper is organized as follows. In Sec.~\ref{sec:exp}, we describe the experimental setup and the system under study.
  In Sec.~\ref{sec:results}, we present experimental data on optical spectra of the SN power within
   the D$_2$ resonance of cesium atoms and results of correlation experiments in the two-color arrangement. 
    In Sec.~\ref{sec:teor}, theoretical description of the obtained experimental results is presented.
      Section~\ref{sec:discussion} discusses our findings.
The concluding remarks are made in Sec.~\ref{sec:concl}.

\section{Sample and setup}
\label{sec:exp}

\begin{figure}[t]
\includegraphics[width=0.75\linewidth]{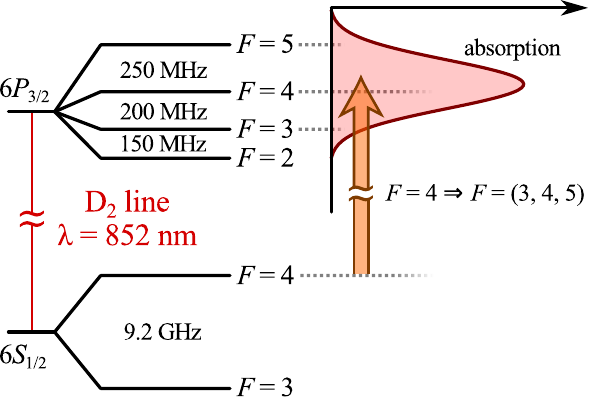}
\caption{
 Energy-level diagram relevant to D$_2$  line of cesium atom. Right vertical arrow indicates the probed transition.
}\label{fig:1}
\end{figure}

The measurements of the SN spectra were performed on D${_2}$ line of cesium atoms ($\lambda = 852$~nm), corresponding to the transition $6S_{1/2}$ ($F = 3,4$) $\leftrightarrow$ $6P_{3/2}$ ($F =2 \ldots 5$), see Refs.~\cite{Chalupczak,cesiumDline} and Fig.~\ref{fig:1}. 
Here $\bm F = \bm L + \bm S+ \bm I$ is the total angular momentum  of the atom  formed by   the electron orbital angular momentum $\bm L$, electron spin $\bm S$ and  nuclear spin~$\bm I$. 
Hyperfine splitting of the excited state is smaller than Doppler broadening of the transition at  room temperature (which is about $400$ MHz) and, therefore, is not resolved in the absorption spectrum. 
So, the absorption line D${_2}$ consists of two slightly asymmetric components resulting from hyperfine splitting  of the ground state into sublevels $F=3,4$ ($9.2$~GHz). 
Our measurements were mainly performed on the transition from the ground-state sublevel $F = 4$ (indicated by the right vertical arrow in Fig.~\ref{fig:1}). 

The cell with cesium vapor, $\sim$$20$~mm in length, contained a buffer gas (neon) under pressure of 1 Torr and was held at a fixed temperature in the range of $35$--$50$~\textdegree{}C. It is important to note that the collisional broadening of the transition, under these conditions, was ${\nu_c}\sim 10$~MHz (free path length ${\lambda}\sim 50$~\textmu{}m) is small, and the total width of the transition (around $800$~MHz) resulting from combined action of Doppler broadening and hyperfine splitting of the excited state could be definitely considered as inhomogeneous (see Sec.~\ref{sec:results}).

\begin{figure}
\includegraphics[width=1\linewidth]{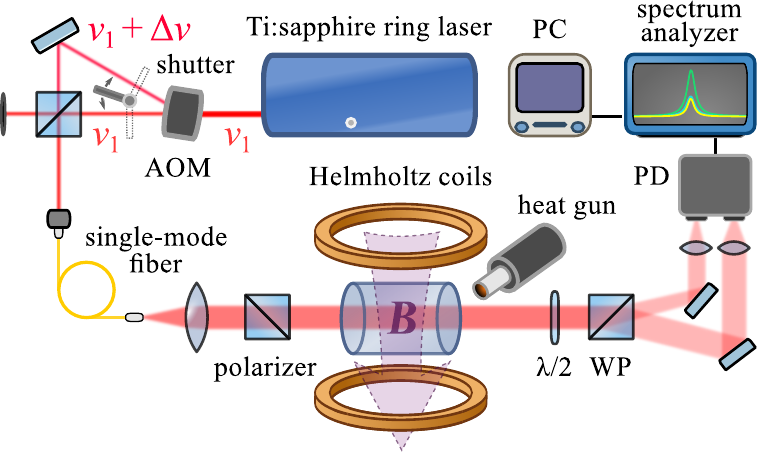}
\caption{Schematic of the experimental setup.  Output emission of the frequency-stabilized  Ti:sapphire ring laser is split by the acousto-optical modulator (AOM) to obtain two beams with the frequencies $\nu_1$ and $\nu_2$ shifted by 250 MHz.   The beams  are transmitted through the single-mode fiber and are used (either alternatively or together)  to probe  the Cs cell.  WP --  Wollaston prism, PD -- balanced photodetector.
}\label{fig:2} 
\end{figure}
 
We used a conventional SNS experimental setup (Fig.~\ref{fig:2}) with some modifications suitable for specific tasks of this work. 
The cell was probed with the linearly polarized light beam of a single-frequency Ti:Sapphire laser whose lasing wavelength was stabilized with a reference cavity and continuously monitored by a Fizeau interferometric device. 
The laser system was operated in a regime that provided the spectral linewidth of the probing beam~$<$$10$~kHz. 
The light was delivered to the cell through a single-mode optical fiber, which was especially convenient in the two-color experiments, when the cell had to be probed simultaneously with two light beams of different wavelengths with identical spatial characteristics. 
Polarization noise of the beam transmitted through the cell was detected with a balanced photoreceiver and transformed into spectral domain with a FFT spectrum analyzer. 
Optical frequency of the output laser beam could be smoothly tuned within the range of a few GHz that substantially exceeded total width of the transition and allowed us to obtain high-resolution spectra of transmission and SN power throughout the D${_2}$ line. 
To minimize the effects of optical perturbation, governed by the probe beam power density in the cell, we employed a collimated probe of a relatively large diameter (\mbox{$4$--$12$}~mm) rather than a focused beam commonly used in the SNS measurements.
 
The measurements were made in the laboratory magnetic field, with its longitudinal (with respect to the light propagation) component compensated with a pair of Helmholtz coils (not shown in Fig.~\ref{fig:2}). 
Uniformity of the field within the probed volume of the cell was good enough to observe a narrow isolated SN resonance ($\delta f \simeq 5$~kHz) at a frequency $f \simeq 160$ kHz corresponding to the value of the Land\'e factor $g = 0.25$ for $B = 0.45$~G, see Fig.~\ref{fig:3}(a).

\section{Results of the measurements}
\label{sec:results}

\subsection{Spin noise spectra}

\begin{figure}
\includegraphics[width=1\linewidth]{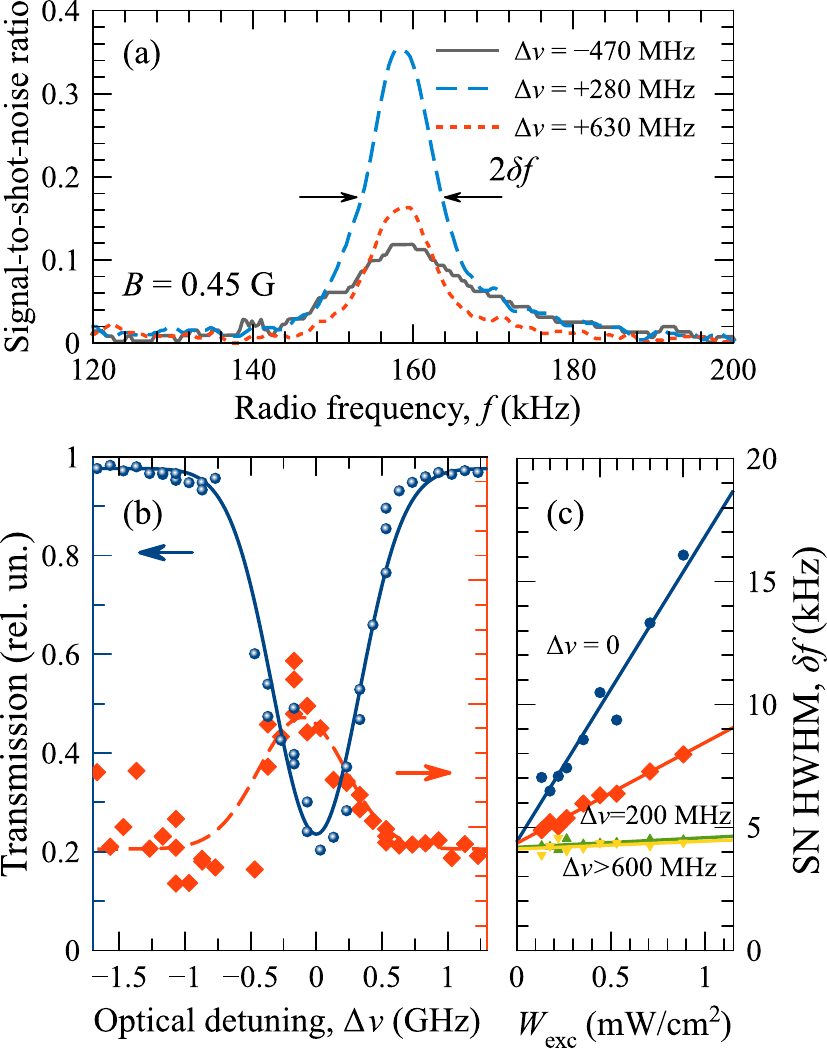}
\caption{
 Behavior of the SN spectra versus optical frequency of the probe beam in the region of the D${_2}$ line of cesium.   (a) SN spectra measured at different detunings from the D${_2}$ absorption resonance.  (b) Optical spectrum of transmission (blue circles) and of SN resonance width (red diamonds) throughout the D${_2}$ resonance. The lines show Gaussian fits for these data ($\text{HWHM} = 390$~MHz). (c) Intensity dependence of the light-induced broadening of the SN resonance at different detunings $\Delta\nu$.  $T = 38$~\textdegree{}C. The probe power density in (a,b) is $W_\mathrm{exc} = 0.47$ mW/cm$^2$.}
\label{fig:3}
\end{figure}

Figure~\ref{fig:3}(a) shows the measured SN spectra for three different values of detuning $\Delta \nu$ from the D${_2}$ resonance ($\nu=351.725$~THz) measured at a fixed probe power density at the laboratory transverse magnetic field of about $0.45$~G. Since the main goal of our work was to study the effect of Doppler broadening  on the SN, we needed to tune the probe into the absorption line, where the probe-induced perturbation of the spin-system was inevitable. To minimize this perturbation and to remain in the framework of linear optics we reduced the light power density and increased the signal accumulation time up to 1--3~min. This allowed us to observe SN spectra like shown in Fig.~\ref{fig:3}(a) for the whole range of detunings under study and to reliably determine their half-width at half-maximum (HWHM) and the area. 

The spectra in Fig.~\ref{fig:3}(a) are in good agreement with previous measurements on this system~\cite{Chalupczak} demonstrating Lorentzian shape. For large detunings between the probe beam frequency and the D${_2}$ line, the SN spectra are rather narrow, with the HWHM $\delta f$ below $5$~kHz. These spectra measured at large detunings are practically unaffected by variations of the probe power density $W_{\rm exc}$, whereas with decreasing detuning, optical perturbation of the spin system becomes significant. 

By measuring spectral dependence of the SN resonance width, we have found that, indeed, it exhibited additional broadening inside the optical transition, where the effect of fluctuating detuning was expected to be most pronounced [Fig.~\ref{fig:3}(b)]. 
This broadening, however, varied with the probe beam intensity [Fig.~\ref{fig:3}(c)] and, therefore, resulted from optical perturbation of the spin-system as, e.g., reported for Cs vapors in Ref.~\onlinecite{Chalupczak}. 
In the range of relatively low light intensities, this probe-induced contribution to the broadening varied linearly with the probe light intensity, and the residual SN resonance width (interpolated to zero intensity) at all wavelengths coincided with that observed outside the optical resonance [Fig.~\ref{fig:3}(c)]. 
It means that, in the framework of linear optics, the SN resonance width, under our experimental conditions, does not reveal any broadening associated with the rapidly fluctuating in time Doppler shifts of optical resonances of the atoms.

\subsection{Optical spectrum of the SN power}

The Doppler width of the D${_2}$ line under our experimental conditions, essentially exceeded its homogeneous width, and in terms of conventional optical spectroscopy the line was broadened inhomogeneously. 
Therefore, it is expected, in accordance with the results of Ref.~\onlinecite{Zapasskii13}, that the optical spectrum of the SN power should not exhibit any dip at the line center, where the FR itself vanishes. 
To check it, we performed spectral measurements of the SN power at the lowest level of the light power density ($W_\mathrm{exc} = 0.25$~mW/cm$^2$) where the effects of optical perturbation of the spin-system could be neglected. 
In these experiments, the diameter of the light beam was increased up to $\sim$$12$~mm to retain optimal level of the light power on the detector at lowest light power density in the cell. 

\begin{figure}
\includegraphics[width=1\linewidth]{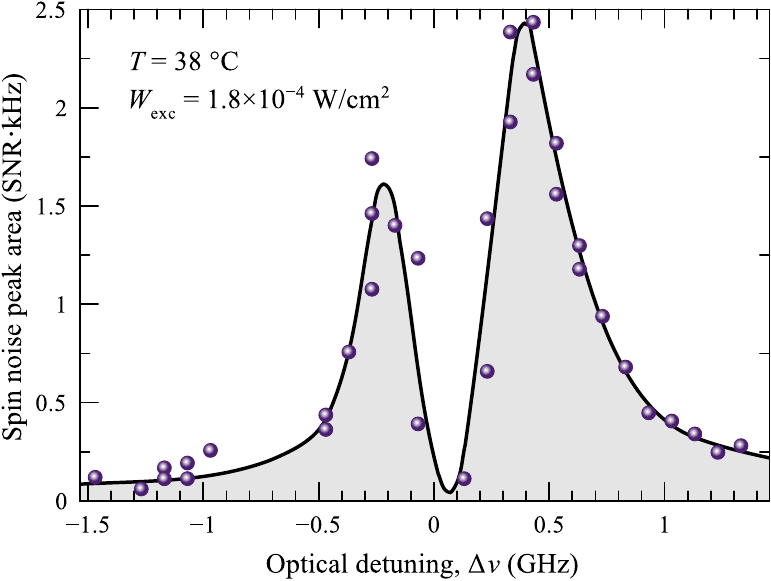}
\caption{
SN power  as a function of  detuning of the probe beam frequency with respect to the D${_2}$ resonance (circles). Solid line is a guide for the eye.
}\label{fig:4}
\end{figure}

Our measurements presented in Fig.~\ref{fig:4} have shown, however, that the observed spectral shape of the SN power, in the region of the optical transition, was typical for the {\it homogeneously broadened} line, with a dip down to zero at the line center. In spite of the fact that the packet of the near-resonant atoms, making main contribution to the detected SN, is refreshed entirely for fractions of microsecond, the detected phase of the precession signal remains undisturbed for hundreds of microseconds. 
It looked like the detected signal was insensitive to fluctuating detuning of individual atoms and perceived the system as a homogeneous ensemble of atoms with some averaged effective detuning. 
In this case, however, spin fluctuations detected at different wavelengths inside the line should be correlated as if the Doppler broadening were homogeneous. This supposition has been checked experimentally. 

\subsection{Correlation of the SN over the Doppler-broadened line.} \label{IIIC}

The two-color experiment is an alternative approach to verify the assumption that Doppler broadening is revealed in SNS as homogeneous. Using an acousto-optical modulator we generated a second beam with a frequency shifted by a value of $\Delta\nu= 250$~MHz ($\nu_2 = \nu_1+\Delta \nu$) that strongly exceeded the homogeneous (collisional) width of the optical transition ($\sim$$10$ MHz). Therefore, we could suppose that, under resonant probing, these two beams would monitor different, independently fluctuating spectral packets of the inhomogeneously broadened line. 
As shown in Fig.~\ref{fig:2}, the two beams follow the same optical path, and the SN spectrum can be obtained using separately each of them or both. It is clear that, when two beams are combined, their fluctuations will sum up differently depending on whether they are correlated or not. By comparing the noise power for each of the beams with that for their mixture, we can judge about degree of correlation of fluctuations in 
the two beams. This is the basic idea of our simple correlation experiment. 

In our case, it is useful to distinguish two situations, with signs of detuning for the two beams being (a) different or (b) the same.
These situations are essentially different because the FR spectrum of a spin polarized system is an odd function of the detuning (see red curve in Fig.~\ref{fig:theory} and discussion in Sec.~\ref{sec:teor} below). For the homogeneously broadened line, fluctuations of the FR are anticorrelated in case (a) and correlated  in case (b) (positive fluctuation of FR on one side of the line is accompanied by negative fluctuation on the other).

\begin{figure}
\includegraphics[width=1\linewidth]{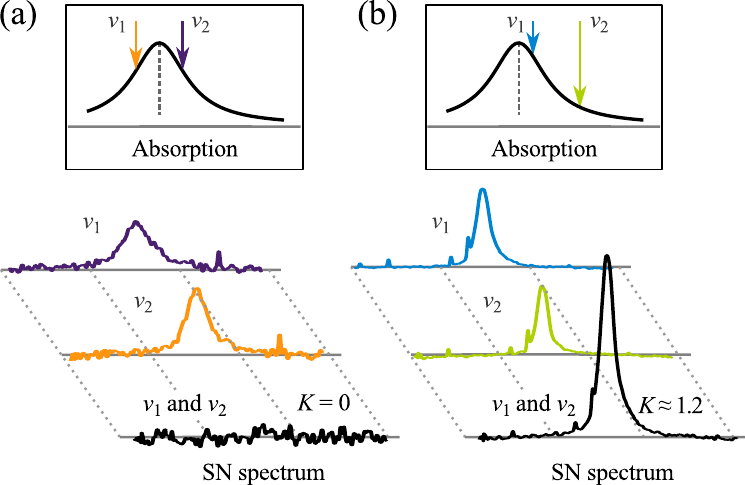}
\caption{
Two-color SN spectra (lower panels) measured with the probe frequencies $\nu_1$ and $\nu_2 = \nu_1+250$ MHz lying (a) on the opposite sides of the D${_2}$ resonance and (b) at the same side of the resonance. Spectral positions of the beams with respect to the line profile are indicated by arrows on the upper  panels. As seen from the spectra, in the case (a), the two spectra detected at frequencies $\nu_1$ and $\nu_2$, being detected  in the two-frequency beam ($\nu_1$ and $\nu_2$), cancel each other, while in the case (b), they enhance each other. In both cases, the light power density created by a single-frequency beam was $W_\mathrm{exc} = 1.8 \cdot 10^{-4}$~W/cm$^2$. }
\label{fig:5}
\end{figure}

Results of the two-color SNS for our system are shown in Fig.~\ref{fig:5}. We have found that intensity of the SN spectrum detected in the two-frequency beam ($\nu_1 $and $ \nu_2$), $P_{\rm both}$, crucially depended on mutual position of the frequencies $\nu_1$ and $\nu_2$ with respect to the line center. In the case (a), when signs of detuning for the two frequencies were different, the two spectra obtained with each of the beams cancelled each other, and the SN power vanished [Fig.~\ref{fig:5}(a)], $P_{\rm both}=0$, while in the case (b), when two frequencies lay on the same side of the line, the two spectra combined constructively and the SN resonance was significantly 
enhanced [Fig.~\ref{fig:5}(b)]. We introduce the correlation coefficient
  \begin{equation}
    \label{eq:corr}
    K = P_\mathrm{both}/(P_1 + P_2),
  \end{equation}
where $P_1$ and $P_2$ are the intensities of the SN in the individual beams with the frequencies $\nu_1$ and $\nu_2$, respectively. In our case it equals to $K\approx 1.2$, which is larger than $1$ expected for the non-correlated spin fluctuations.

These results strongly support our assumption about total correlation of the optically detected spin fluctuations over the Doppler-broadened line. Specifically, our experimental data show that by changing the probe beam wavelength within the Doppler width of the atomic resonance, we do not select different sub-ensembles of atoms with independently fluctuating magnetizations, as could be expected by analogy with inhomogeneously broadened ensembles of quantum dots~\cite{Zapasskii13,Yang:2014aa}. To figure out the reason of this result, which may seem paradoxical, we have undertaken the detailed theoretical analysis of the problem.

\section{Theory}
\label{sec:teor}

In this section we discuss mechanisms of the FR fluctuations, broadening of the transmission spectrum and optical SN spectrum, as well as the correlation of the FR noise detected in the two-color configuration.

We consider a gas of atoms with optical resonance at the frequency $\omega_0$. We recall that the motion of the atom with the velocity $\bm v$ results in the Doppler shift of its observed resonance frequency as
\begin{equation}
\label{Doppler}
 \omega(\bm v)=\left(1+\frac{v_z}{c}\right)\omega_0,
\end{equation}
where $z$ is the axis of light propagation, $c$ is the speed of light, and it is assumed that the atomic motion is nonrelativistic, $v/c\ll 1$. Particularly, the secondary electromagnetic field scattered by the atom with the velocity $\bm v$ in the direction of the $z$-axis, contains a resonant contribution~\cite{fabelinskii,gorb_perel,NonlinearSNS}
\begin{equation}
\label{scatter}
\bm E_1 = e^{-\mathrm i \omega t + \mathrm i q z} \frac{\mathrm i \alpha \bm E_0 + \beta \bm E_0 \times \bm e_z {F_z} }{\omega(\bm v) - \omega - \mathrm i \gamma_0}.
\end{equation}
Here $\alpha$ and $\beta$ are real coefficients describing the strength of the light-matter coupling, the contribution $\propto \alpha$ is responsible for the light scattering by an unpolarized atom and the contribution $\propto \beta$ is responsible for the FR due to the spin component $F_z$ of the atom, $\gamma_0$ is the homogeneous width of the atomic resonance, $\omega$ is the light frequency, $q=\omega/c$ is the light wavevector, and $\bm e_z$ is the unit vector along $z$ direction. For unpolarized atoms in thermal equilibrium, the intensity transmission coefficient, $\mathcal T$, is related to the first contribution in Eq.~\eqref{scatter}, and can be written in the form
\begin{equation}
\label{T:I}
\mathcal T = 1 - \int d\bm v \frac{2\alpha N\gamma_0 f(\bm v)}{[\omega(\bm v) - \omega]^2+ \gamma_0^2}.
\end{equation}
In Eq.~\eqref{T:I} $N$ is the number of atoms in the illuminated part of the gas, $f(\bm v)$ is the Maxwell-Boltzmann distribution of the atomic velocities,
\begin{equation}
 f(\bm v)=\frac{1}{\pi^{3/2}v_T^3}\e^{-v^2/v_T^2}, \quad \int d\bm v f(\bm v) =1,
\end{equation}
where
\begin{equation}
 v_T=\sqrt{\frac{2k_B T}{m}}
\end{equation}
is the thermal velocity, with $T$ being the temperature, $k_B$, the Boltzmann constant, and $m$, the atom mass. Equation~\eqref{T:I} holds for not too dense gas where the multiple scattering of light by the atoms can be disregarded.

Thermal motion of the atoms results in broadening of the resonance in the transmission spectrum due to the Doppler effect, Eq.~\eqref{Doppler}. Provided that $\gamma\equiv \omega_0 v_T/c \gg \gamma_0$, the transmission is described by the following simple expression,
\begin{equation}
\label{T:I:1}
\mathcal T = 1 - \frac{2\sqrt{\pi} \alpha N}{\gamma} \exp{(-\Delta^2/\gamma^2)},
\end{equation}
where we have introduced the detuning $\Delta=\omega-\omega_0=2\pi\Delta\nu$. Equation~\eqref{T:I:1} has a characteristic Gaussian form and is typical for the inhomogeneous broadening of the spectrum. This expression is shown by the blue line in Fig.~\ref{fig:3} and describes fairly well the transmission spectrum of the gas. 

\begin{figure}[t]
\includegraphics[width=1\linewidth]{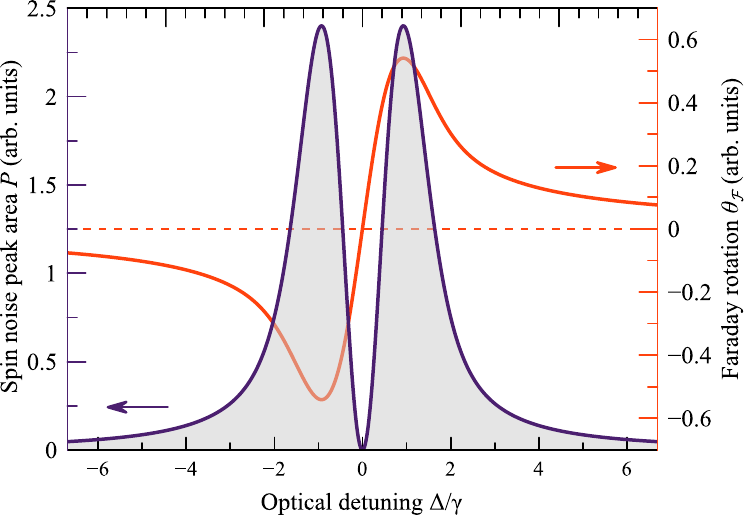}
\caption{The Faraday rotation spectrum $\theta_{\mathcal F}$ calculated after Eq.~\eqref{static:F} (red line) and spin noise power spectrum $P$ calculated after Eq.~\eqref{eq:D2} (violet line).}
\label{fig:theory}	
\end{figure}

Now we turn to the description of the static and fluctuating FR in the atomic gas. In order to calculate the static FR, let us assume that all atoms are polarized (i.e., by an optical pumping or by a strong static magnetic field), ${\bm F}\parallel z$. The static FR angle can be evaluated following Refs.~\onlinecite{gorb_perel,NonlinearSNS,Ma:2017aa} with the result
\begin{equation}
\label{static:F}
\vartheta_{\mathcal F} = \beta N {F_z} \int d\bm v \frac{f(\bm v) [\omega(\bm v) - \omega]}{[\omega(\bm v) - \omega]^2+ \gamma_0^2} \approx \frac{2\beta N {F_z}}{\gamma}D(\Delta/\gamma),
\end{equation} 
Here we neglected the difference between $\mathcal T$ and unity, $D(x) = \exp(-x^2) \int_0^x \exp(y^2) dy$ is the Dawson integral. The last approximate equality in Eq.~\eqref{static:F} holds for $\gamma \gg \gamma_0$. Equation~\eqref{static:F} describes the profile of the FR for inhomogeneous ensemble of polarized spins with zero, as usual, exactly in the resonance, $\Delta=0$, and opposite signs of the rotation for positive and negative detunings, see red curve in Fig.~\ref{fig:theory}. Noteworthy, Eq.~\eqref{static:F} describes the conversion of spin polarization into the FR both in atomic gas and in quantum dot ensembles~\cite{glazov:review}.

Our next goal is to calculate the spectrum of the FR fluctuations in the unpolarized gas. Due to the fact that in atomic vapors the spin relaxation time $\tau_s$ exceeds by far the atom scattering (momentum relaxation) time $\tau_p=1/(2\pi\nu_c)$, with $\nu_c$ being the collision frequency, the atom experiences many collisions before its spin fluctuation vanishes. Since at each collision the atomic velocity changes randomly, its Doppler shift changes. By contrast, the spin fluctuations of individual atoms are uncorrelated. In order to evaluate the FR fluctuations, we present the contribution of the individual atom to the FR at a time $t$ as
\begin{equation}
\label{single}
\delta \vartheta_{\mathcal F}(t) = \beta {F_z}(t) \mathcal F(t),
\end{equation} 
with ${F_z}$ being the momentary $z$-spin component of the atom
\begin{equation}
 \mathcal F(t)=\frac{\omega-\omega[\bm v(t)]}{\left\{\omega[\bm v(t)]-\omega\right\}^2+\gamma_0^2},
 \label{eq:F}
\end{equation}
representing the profile of the FR for an atom moving with the velocity $\bm v$. In what follows, we assume that the light propagation time through the sample is much shorter than both $\tau_p$ and $\tau_s$, and that the spin and kinetic dynamics of the atom are uncorrelated. The latter assumption is justified by the fact that $\tau_s \gg \tau_p$, moreover, in the experimental conditions the effective spin relaxation can be related to the atoms leaving the illuminated beam. Under this assumption the correlation function of FR takes the form
\begin{equation}
 \aver{ \vartheta_{\mathcal F}(t) \vartheta_{\mathcal F}(t+\tau)}=\beta^2 N\aver{{F_z}(t){F_z}(t+\tau)}\aver{\mathcal F(t)\mathcal F(t+\tau)}.
\end{equation}
Here the angular brackets denote ensemble averaging and averaging over the time $t$ at a fixed delay $\tau$. Since the average $\aver{\mathcal F}$ is nonzero at $\Delta \ne 0$, it is convenient to separate two contributions to the FR correlator as
\begin{widetext}
\begin{equation}
 \aver{ \vartheta_{\mathcal F}(t) \vartheta_{\mathcal F}(t+\tau)}=\beta^2 N\aver{{F_z}(t){F_z}(t+\tau)}\left(\aver{\mathcal F}^2+\aver{\delta \mathcal F(t)\delta\mathcal F(t+\tau)}\right),
\label{eq:2contrib}
\end{equation}
\end{widetext}
where $\delta\mathcal F(t)=\mathcal F(t)-\aver{\mathcal F}$. The both contributions are proportional to the spin correlation function, which decays to zero during the characteristic time $\tau_s$. The second term contains additional information about velocity dynamics, and decays to zero during the time $\tau_p$.

The FR noise spectrum as a function of low (noise) frequency $\Omega=2\pi f$ is defined as
\begin{equation}
\label{Omega:corr}
 (\vartheta_{\mathcal F}^2)_\Omega=\int\limits_{-\infty}^\infty\aver{\vartheta_{\mathcal F}(t)\vartheta_{\mathcal F}(t+\tau)}\e^{\i\Omega\tau}\d\tau.
\end{equation}
We recall that the correlation function is an even function of $\tau$~\cite{ll5_eng}, and accordingly the SN spectrum is an even function of $\Omega$. Experimentally the spin noise power is measured as an area under the spectrum:
\begin{equation}
\label{power}
 P=\int\limits_0^{\Omega_{\rm max}}(\vartheta_{\mathcal F}^2)_\Omega\d\Omega,
\end{equation}
where $\Omega_{\rm max}$ is the upper limit of frequency bandwidth for spin noise measurements. In atomic gas the limit $\tau_p\ll\tau_s$ is realized. In this case the second contribution in Eq.~\eqref{eq:2contrib} rapidly decays to zero on the timescale $\sim \tau_p$, while the spin fluctuation on this timescale can be considered as frozen. Accordingly the FR noise spectrum consists of two contributions:
\begin{equation}
 (\vartheta_{\mathcal F}^2)_\Omega=\beta^2 N\aver{\mathcal F}^2({F_z^2})_\Omega+\beta^2 N (\delta\mathcal F^2)_\Omega.
 \label{eq:sns}
\end{equation}
The noise powers $({F_z^2})_\Omega$ and $(\delta\mathcal F^2)_\Omega$ are defined analogously to Eq.~\eqref{Omega:corr}.
In our experiment $\tau_p$ is of the order of $16$~ns, so the second contribution in Eq.~\eqref{eq:sns} is beyond the detector bandwidth ($\Omega_{\rm max}=600$~kHz): $\Omega_{\rm max}\tau_p\ll 1$. Therefore only the first term in Eq.~\eqref{eq:sns} contributes to the measured spin noise power, and we obtain its optical spectrum
\begin{equation}
 P \propto \frac{\beta^2N}{\gamma^2}D^2\left(\frac{\Delta}{\gamma}\right).
\label{eq:D2}
\end{equation}
This expression is in agreement with Eq.~\eqref{static:F}. Indeed, during the spin relaxation time $\tau_s$ any atom experienced many collisions and, hence, ``probed'' all possible detunings caused by the Doppler shift [Eq.~\eqref{Doppler}]. Thus, in order to calculate the SN power one can apply Eq.~\eqref{static:F} assuming that ${F_z}$ is the total spin fluctuation.

The SN power spectrum is shown by the blue line in Fig.~\ref{fig:theory} and demonstrates a characteristic dip down to $0$ at the resonance, $\Delta=0$, in good agreement with experiment. This dip is a result of the interference of the contributions from the same atom (and, hence, same spin fluctuations) at different time moments where its instantaneous detuning from the resonance, $\omega(\bm v) - \omega$, was positive and negative.

Such a behavior of the spin noise power is strongly different from that observed in inhomogeneously broadened quantum dot ensembles~\cite{Zapasskii13}. In quantum dot structures the detuning of the quantum dot is determined by its geometrical parameters and composition, it does not fluctuate in time. The power spectrum of the spin noise in this case can be formally obtained from Eqs.~\eqref{eq:2contrib} and \eqref{power} by considering the limit of $\tau_p \gg\tau_s$ as
\begin{equation}
 P \propto 
 \frac{\beta^2N}{2\gamma\gamma_0}\e^{-\Delta^2/\gamma^2},
 \label{eq:P_QD}
\end{equation}
in agreement with previous works~\cite{gi2012noise,Zapasskii13}. One can see that this spectrum has the maximum at zero detuning, $\Delta=0$.

Let us now consider the two-color experiment, when the atomic gas is probed by the light with carrier frequencies $\omega_1 = 2\pi \nu_1$ and $\omega_2= 2\pi \nu_2$ (with corresponding detunings $\Delta_{1,2}=\omega_{1,2}-\omega_0$). In this case, the above analysis remains valid with the replacement
\begin{equation}
 \beta\mathcal F\to\beta(\mathcal F_1+\mathcal F_2),
 \label{eq:rep}
\end{equation}
where $\mathcal F_{1,2}$ stands for Eq.~\eqref{eq:F} with $\omega=\omega_{1,2}$, respectively, and we assumed equal intensities of the probe beams, as used in the experiment.

The SN power, as introduced in Eq.~\eqref{power} has, under the condition $\tau_s \gg\tau_p$, the form
\begin{equation}
 P\propto \beta^2 N\left(\aver{\mathcal F_1}+\aver{\mathcal F_2}\right)^2.
 \label{eq:P_2color}
\end{equation}
The correlation coefficient calculated after Eq.~\eqref{eq:corr} reads
  \begin{equation}
    K=1+\frac{2\aver{\mathcal F_1}\aver{\mathcal F_2}}{\aver{\mathcal F_1}^2+\aver{\mathcal F_2}^2}.
    \label{eq:K2}
  \end{equation}
Equation~\eqref{eq:K2} clearly demonstrates the specific correlation properties of the SN signal detected by the two-color experiment. Since the signs of $\aver{\mathcal F_{1,2}}$ coincide with the signs of $\Delta_{1,2}$, the detected fluctuations are correlated if the probe beams have the same sign of the detuning ($K>1$), $\sign{(\Delta_1\Delta_2)}>0$, and are anticorrelated otherwise. Particularly, if detunings are chosen in such a way that $\aver{\mathcal F_{1}}^2=\aver{\mathcal F_{2}}^2$ the correlation coefficient
\begin{equation}
 K=1+\sign{(\Delta_1\Delta_2)}.
 \label{eq:P4}
\end{equation} 
As a result, the total SN power $P_{\rm both}$ can be either twice larger than $P_1+P_2$ or zero, depending on the relative signs of $\Delta_1$ and $\Delta_2$.
This behavior is in agreement with experiment as shown in Fig.~\ref{fig:5}. Somewhat smaller experimental value of the correlation coefficient than $2$ ($K\approx 1.2$), is, most probably, related to  perturbation of atomic resonances by the probe beams and to distinction between the values $\mathcal F_1$ and $\mathcal F_2$  at the wing of the line [see Eq.~\eqref{eq:K2}].  Similar behavior of the SN should be evidently expected for a homogeneously broadened resonance because the FR signals at  positive and negative detunings have opposite signs. For completeness, note that the ellipticity noise signal related to a difference in absorption of circularly polarized components is expected to demonstrate positive correlations regardless of the signs of detuning, because the ellipticity signal is an even function of the detuning.

Again, this result  strongly differs from the situation observed in inhomogeneous quantum dots ensembles. There, the correlations are absent provided that the difference of the probe frequencies exceeds the homogeneous linewidth, $|\Delta_1-\Delta_2|\gg \gamma_0$~\cite{Yang:2014aa}. Indeed, formally putting $\tau_p \gg \tau_s$ and making use of Eqs.~\eqref{eq:P_QD} and \eqref{eq:rep} we obtain
\begin{equation}
 P \propto \beta^2 N\left(\aver{\mathcal F_1^2}+\aver{\mathcal F_2^2}\right),
 \quad
K=1,
\end{equation}
i.e., any correlations are absent. Only if $|\Delta_1-\Delta_2|\lesssim \gamma_0$, the correlation in the two-color experiment on quantum dot ensemble appears, $\aver{\mathcal F_1\mathcal F_2} \ne 0$. It has made possible to evaluate the homogeneous linewidth of the quantum dot resonance from the spin noise spectroscopy in Ref.~\cite{Yang:2014aa}.

\begin{figure}
\includegraphics[width=1\linewidth]{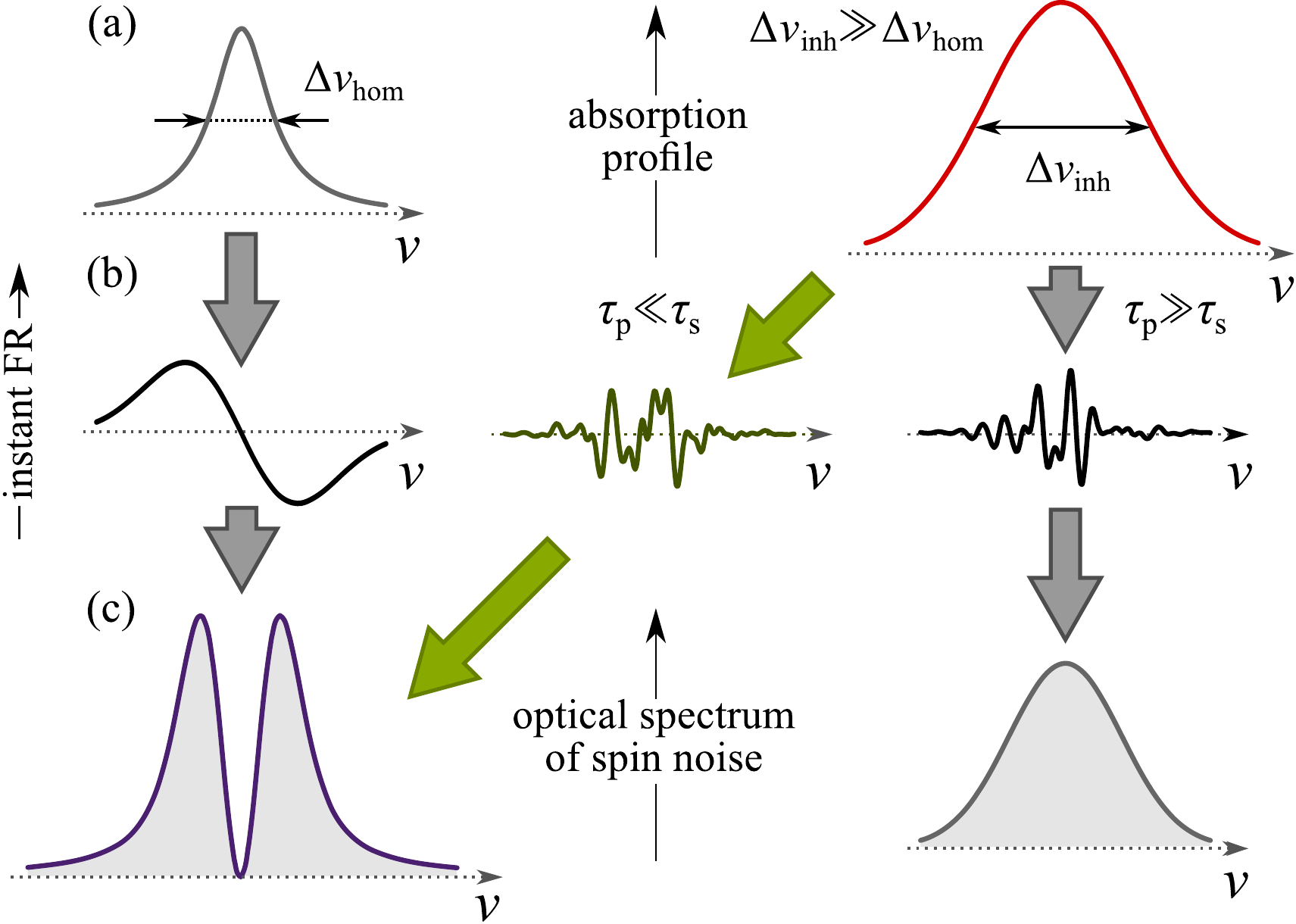}
\caption{
Behavior of the FR noise spectra for optical transitions with different types of broadening.  (a) Absorption spectra of the homogeneously  (left) and inhomogeneously (right) broadened  lines. (b) Instantaneous random realizations of the FR spectrum for homogeneously broadened line (left) and for inhomogeneously broadened line with short (middle) and long (right) $\tau_p$.  Spectral correlation length of the latter spectra reflects   homogeneous width of the optical transitions.  (c) Optical SN spectra for the above  cases of line broadening.}
\label{fig:6}	
\end{figure}

\section{DISCUSSION}
\label{sec:discussion}
 
In Ref.~\onlinecite{Zapasskii13}, there have been considered two types of the optical SN spectra corresponding to two ``pure'' types of optical transition broadening: homogeneous and inhomogeneous (see Fig.~\ref{fig:6}). The strong qualitative difference between the corresponding optical spectra of SN power has been established. 
In this paper, we show that there exists one more type of broadening---Doppler broadening in atomic ensembles with relatively fast momentum relaxation, $\tau_p \ll \tau_s$ that, being fundamentally inhomogeneous, manifests itself in SNS as homogeneous. To make the distinction between these three situations more visual, it is helpful to consider  {\it instantaneous optical spectra} of the fluctuating FR. These are the dependencies of the FR angle on the probe frequency for a given time moment. These spectra fluctuate with time.

For the ``purely homogeneously'' broadened line, the optical spectrum of the ensemble reproduces that of a single atom. In this case the instantaneous optical spectra of the FR for different time moments differ only in amplitudes and signs, but not in shape [Fig.~\ref{fig:6}(b), left spectrum]. In case of inhomogeneous broadening each atom provides its own contribution with small homogeneous width centered at its resonant frequency. As a result, fluctuations of the FR at different optical frequencies (spaced by more than homogeneous width) appear to be uncorrelated, and the instantaneous FR spectrum represents a spectral noise with the correlation length corresponding to homogeneous width of the transition [Fig.~\ref{fig:6}(b), two right spectra].
The difference between static inhomogeneous broadening [denoted by $\tau_p \gg \tau_s$ in Fig.~\ref{fig:6}] and Doppler-induced inhomogeneous broadening is revealed in time evolution of instantaneous optical spectra.
In the latter case of $\tau_p \ll \tau_s$,   the correlation between different frequencies in the optical spectrum arises, at the time-scale $\sim\tau_p$,  due to changes of the resonant frequency of each atom. This is, in essence, the mechanism of ``homogenisation'' of the optical  SN spectra in the Doppler-broadened systems. As a result, the optical SN spectra  in case of Doppler broadening appear to be similar to those for  homogeneously broadened lines [left curve in Fig.~\ref{fig:6}(c)]. 
This explains why the Doppler-broadened optical SN spectrum reveals a dip in the line center that makes it similar with that of the homogeneously broadened line.  We stress, however, that  similarity of these spectra does not imply similarity of the FR noise patterns. As seen from instantaneous FR spectra in Fig.~\ref{fig:6}, in  the case of real homogeneous broadening, fluctuations of the FR in the line center are absent indeed, while in the case of inhomogeneous broadening, they are the greatest but, because of their fast dynamics [second contribution in Eq.~\eqref{eq:sns}], can not be observed in the detected frequency range.

One can readily imagine that, for the wavelength of the probe beam at the line center, total contribution of each atom to the FR signal will vanish after averaging  over the time interval $\tau_p<\Delta t<\tau_s$, because its contributions with opposite detunings will cancel each other.
This qualitatively explains the dip in the optical SN spectrum in this case.

For comparison, in the case of static inhomogeneous broadening, the optical SN spectrum does not reveal any dip  at zero detuning, as shown by the right curve in Fig.~\ref{fig:6}(c). In this case, spectral noise of inhomogeneously broadened bands [either in absorption or in refraction, or in FR as shown in Fig.~\ref{fig:6}(b)] can be used to measure homogeneous width of the transition~\cite{zapasskii-corr-anal-spectral-fluc01}. This situation is realized in quantum dot structures~\cite{Zapasskii13,Yang:2014aa} and in the collisionless atomic gases~\cite{Ma:2017aa}.
 
\section{Conclusion}
\label{sec:concl}

To summarize, we have studied the spin noise (SN) of cesium atoms under conditions of resonant probing in the region of D${_2}$ line. 
In terms of conventional optical spectroscopy, the line is broadened inhomogeneously due to the Doppler effect, while the optical spectrum of SN is typical for the line broadened homogeneously. 
In particular, this spectrum displayed the dip at the line center usually regarded as a direct evidence of homogeneous broadening. 
The two-color experiment performed with the two probes spectrally separated far beyond the homogeneous linewidth, revealed total correlation or anticorrelation of the noise signals of the two probes depending on signs of the detuning. We show that these effects stem from homogenization of Doppler broadening due to fast momentum relaxation of atoms as compared with the spin relaxation rate.

Interestingly, under resonant probing of atomic ensemble, the SN spectrum may remain practically unaffected by the fast fluctuations of   atomic resonance frequencies. 
Generally, this is not the case, and the Doppler effect can become noticeable in the spin-noise spectrum when the SN resonance width becomes comparable with the homogeneous (collisional) width of the optical transition. 
In this case, the SNS can be used for studying collisional dynamics of the atomic system. Using the probe beam strictly resonant with the optical transition, one can get rid entirely of contribution of this particular transition to the SN spectrum and detect spin noise of other populated but nonperturbed states provided by other (nonresonant) transitions of the atomic ensemble.

Note, in conclusion, that results of this work discover new potentialities of the spin noise spectroscopy generally inaccessible to linear optics ~\cite{NonlinearSNS}. The discovered ability of SNS not only to penetrate into hidden structure of optical transitions, but also to get specific information about dynamics of randomly moving spins, not revealed explicitly  in optical spectra of the system, considerably  enriches image of this unique method of research and attracts additional attention to informative content of spontaneous fluctuations  as compared with that of linear susceptibility.
\linebreak

\acknowledgements

We highly appreciate support of the Russian Science Foundation (grant No. 17-12-01124). M.M.G. was partially supported by the SPbU and DFG, project No. 40.65.62.2017 and
RF President Grant No. MD-1555.2017.2. D.S.S. was partially supported by the Russian Foundation for Basic Research (Grant No. 17-02-0383), RF President Grant SP-643.2015.5
and Basis Foundation. The work was performed using equipment of the SPbU Resource Center ``Nanophotonics''.

\end{document}